# ProGroTrack: Deep Learning-Assisted Tracking of Intracellular Protein Growth Dynamics

Kai San Chan[1,2,3], Huimiao Chen[1,3], Chenyu Jin[1], Yuxuan Tian[1], Dingchang Lin[2]
Department of Biomedical Engineering, Johns Hopkins University, Baltimore, USA
Department of Material Science and Engineering, Johns Hopkins University, Baltimore, USA
Department of Biomedical Engineering, Tsinghua University, Beijing, China

*Abstract*—**Accurate tracking of cellular and subcellular structures, along with their dynamics, plays a pivotal role in understanding the underlying mechanisms of biological systems. This paper presents a novel approach, *ProGroTrack*, that combines the You Only Look Once (YOLO) and ByteTrack algorithms within the detection-based tracking (DBT) framework to track intracellular protein nanostructures. Focusing on iPAK4 protein fibers as a representative case study, we conducted a comprehensive evaluation of YOLOv5 and YOLOv8 models, revealing the superior performance of YOLOv5 on our dataset. Notably, YOLOv5x achieved an impressive mAP50 of 0.839 and F-score of 0.819. To further optimize detection capabilities, we incorporated semi-supervised learning for model improvement, resulting in enhanced performances in all metrics. Subsequently, we successfully applied our approach to track the growth behavior of iPAK4 protein fibers, revealing their two distinct growth phases consistent with a previously reported kinetic model. This research showcases the promising potential of our approach, extending beyond iPAK4 fibers. It also offers a significant advancement in precise tracking of dynamic processes in live cells, and fostering new avenues for biomedical research.**

## I. Introduction

Tracking the dynamics of biomolecules and intracellular compartments such as proteins and organelles is crucial for understanding the long-term physiological and pathological mechanisms in biological systems [1]. While live-cell imaging techniques, such as time-lapse imaging and other longitudinal examinations [2], enable the capture of dynamic processes and spatiotemporal patterning in living organisms, the volume of data collected exceeds the analytical capacity of human observers. Manually tracking hundreds to thousands of cells or molecules through astronomical amount of image frames is practically infeasible [3]. It is not only labor-intensive and time-consuming, but also susceptible to human error and bias. Therefore, more efficient and accurate image analysis techniques are needed for tracking cellular and molecular activities.

Recent advancements in deep learning-based computational approaches offer the potential for efficient and accurate analysis of high-throughput data [4]. These approaches have been extensively validated across multiple domains [5-7], and various applications have been implemented to support laboratory research and clinical diagnosis in the realm of biomedicine, such as medical image super-resolution [8], clinical lesion detection [9], and cell segmentations [10]. In a contrary, tools for tracking at the cellular and molecular level are limited compared to other deep learning-based computer vision regimes, due to the intrinsic challenges posed by biomedical molecules, such as their structural complexity, low contrast and signal-to-noise ratio, motion blur, and scale variation [2].

To address these issues, multiple object tracking (MOT) techniques, commonly used in macro scenarios such as video surveillance [11], automatic vehicle systems [12], and sports game analysis [13], can be extended to the cellular or molecular scale. Despite differences in the objects being tracked, the majority of MOT paradigms share the same two steps: i) target detection (the spatial aspect), in which spots that stand out from the background according to certain criteria are identified and their coordinates estimated in every frame of the image sequences, and ii) target linking (the temporal aspect), in which detected targets are tracked from frame to frame according to their association with certain characteristics such as intersection-over-union (IOU), appearance feature, etc [2]. This type of processing rationale is referred to as detection-based tracking (DBT).

In this work, we propose *ProGroTrack*, a novel approach for cellular and biomolecular level tracking based on the DBT framework by combining the You Only Look Once (YOLO) [14] and ByteTrack algorithms [15]. YOLO is a one-stage object detector that predicts bounding boxes over targets without the need for a region proposal step, making it much faster than two-stage object detectors such as RCNN and Faster R-CNN [16]. Here, the latest YOLOv8 [17] and well-studied YOLOv5 [18] were selected. For ByteTrack, it is an up-to-date, simple, effective, and generic association method developed in 2022. Instead of only tracking the high score detection boxes, it also uses low ones in the matching processes to recover lost tracklets due to occlusion, motion blur, or size

* This work is funded by NIH MIRA Award (1R35GM147274) and Johns Hopkins University

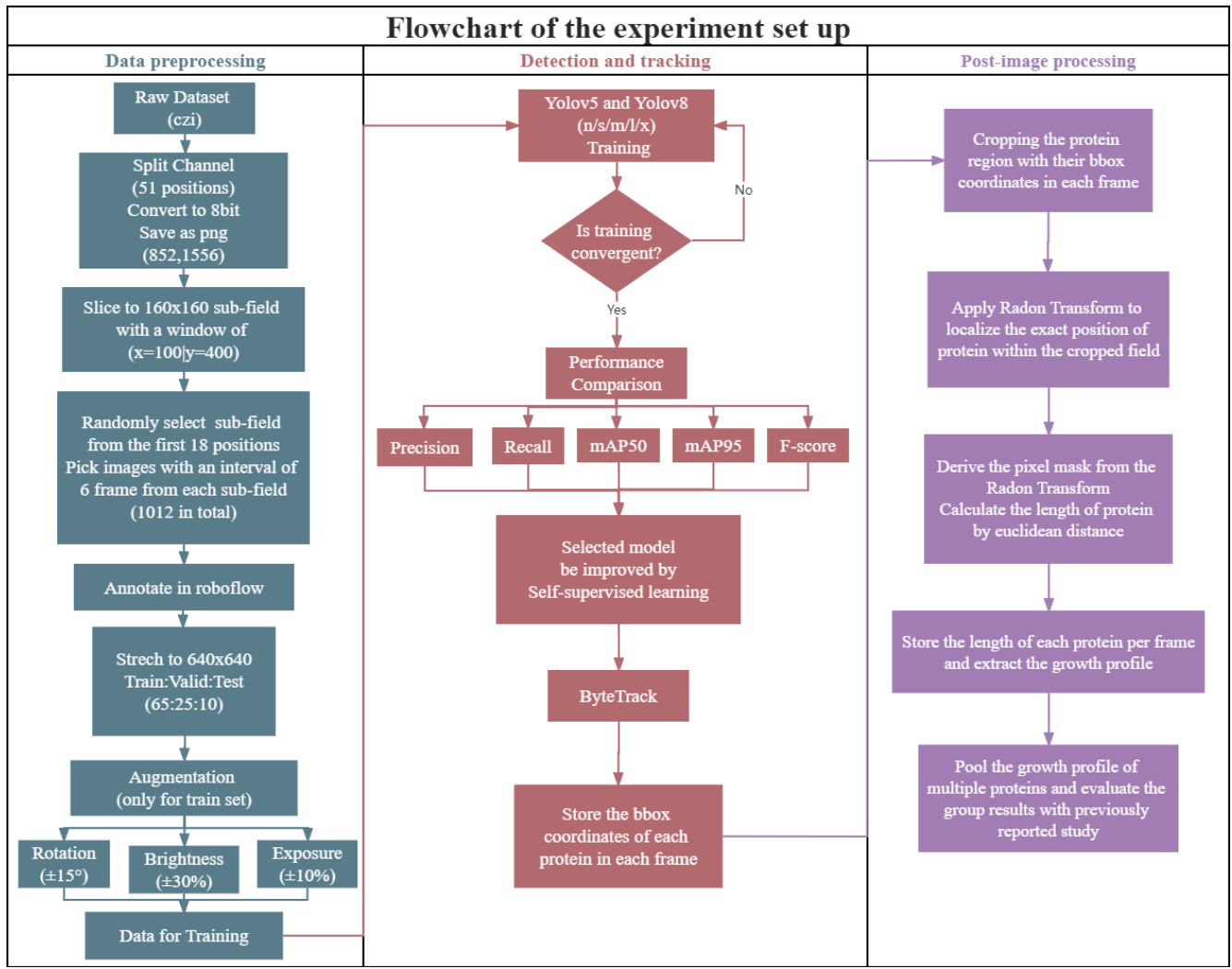

**Figure 1: Flowchart of experimental set up.** Left: The data preprocessing including training data preparation and data augmentation; Middle: The model training and evaluation; Right: The Post-tracking processing including iPAK4 growth file extraction and analysis.

variation. By integrating these two components, our model aims to extend efficient and accurate detection-based tracking to the cellular and molecular level. This paper presents our methodology and experimental setup, along with the results obtained and their related discussion. The potential significance is also highlighted in the end.

## II. METHODOLOGY

The methodology and experimental setup of our work can be categorized into three sections, as depicted in *Figure 1*. This section provides a comprehensive description of our approach. Firstly, we detail the data pre-processing phase, which encompasses the preparation of the training dataset and the implementation of data augmentation techniques. Subsequently, we elucidate the workflow of the model training process, emphasizing the strategies employed to enhance performance. Lastly, we delve into the post-image processing stage, wherein the Radon transform is integrated, and elucidate the extraction methodology for the protein growth profile.

### A. Data Pre-processing

The dataset used in this work was obtained from the previous study which first used engineered iPAK4 protein fibers to store a ticker tape-like transcriptional history of cells [19]. This dataset comprises of time-lapse microscopy images of HEK cells that expressing the engineered iPAK4 protein fibers and cultured on 14-mm glass-bottom dishes These images were captured by Zeiss Elyra8 with 488-nm laser, 1% laser power, 300-ms exposures, and 10-minutes intervals, and were then compressed together and save as a czi video file.

For this paper, a specific video containing time-lapse data from 51 positions (field of view) on the same cell culture dish was selected for model training and demonstration. The video, with a resolution of 852 pixels × 1556 pixels, was initially processed using ImageJ. It was split into two channels, namely the bright field and fluorescent field. The fluorescent field time-lapse for each position was retained, and each frame of the fluorescent field time-lapse was converted to 8-bit and saved as PNG files for further analysis. Since the original resolution was too large for dataset labeling, each frame of each position was subsequently sliced into sub-fields with a resolution of 160 pixels × 160 pixels, using a step size of (x = 100 pixels, y = 400 pixels). As a result, 15 sub-fields could be obtained from each position.

For the training dataset, 24 sliced sub-fields were randomly selected from the first 18 positions. To capture the features of iPAK4 protein fibers during different growth phases, images were picked from each sub-field with an interval of 6 frames (1 hour). After manual filtering, a total of 1012 suitable images that depicted clear and complete protein individuals were intentionally chosen. These images were resized to 640 pixels × 640 pixels to match the input layer of the model. Subsequently, they were uploaded to the free annotation tool, Roboflow [20], for manual annotation of the position and identification of iPAK4 proteins. The annotated figures were divided into training, validation, and test sets in a ratio of 65:25:10. To enhance the diversity and performance of our model, three types of data augmentation techniques were applied to the training set. These techniques included: i) rotation ± 15°, ii) brightness ± 30%, and iii) exposure ± 10%. Rotation randomly rotated images within the range of ± 15°, enabling the model to learn robust features from various orientations and improve its generalization ability. Brightness augmentation altered the pixel intensity within ± 30%, effectively simulating different lighting conditions of the microscope and enhancing the model's adaptability. Furthermore, exposure adjustment simulated diverse lighting environments and improved the model's robustness to changes in illumination. Consequently, there are a total of 1971 images for training, 252 images for validation, and 103 images for testing, while the bounding box annotations of iPAK4 proteins were stored in yaml files for later training.

The dataset for investigating fiber growth profiles was selected separately from other 10 positions to avoid any contamination between training and testing. The sliced sub-fields from these positions were randomly chosen, resized to 640 pixels × 640 pixels, and converted into mp4 video files. These files were utilized for extracting protein growth profiles after completing the model training process.

## B. Detection and Tracking

Model training was carried out following the preparation of the training dataset. In this study, we employed the open-source PyTorch implementation of YOLOv8 [17] and YOLOv5 [18], which were trained on the Google Colab computation platform with an NVIDIA A100 GPU to expedite the training process. Although YOLOv8 and YOLOv5 differ in their configuration, they share the same training strategy. Each model was trained with the previously prepared dataset using a 5-fold cross-validation training process with a batch size of 16 and 100 epochs. The schematic explanation can be referred to as *Figure 2a*. This allowed the model weights to be optimized via backpropagation. After each epoch, the validation set was fed into the model to assess its performance and facilitate hyperparameter tuning. This also helps prevent overfitting and ensures the generalization capabilities of the model. Finally, the test set, which was not used during training and validation, served as the final evaluation target for the trained model.

Upon completion of training, we employed five metrics to quantify the model's performance: mAP50, mAP95, precision, recall, and F-score. mAP50 and mAP95 represent the mean average precision calculated with an IOU threshold of 0.5 and 0.95, respectively, where IOU is the intersection of union between the predicted bounding box and the ground truth. The mathematical formulas for precision, recall, and F-score are as follows:

$$Precision = \frac{True\ Positive}{True\ Positive + False\ Positive} \quad (1)$$

$$Recall = \frac{True\ Positive}{True\ Positive + False\ Negative} \quad (2)$$

$$F\ score = \frac{2 \times (Precision \times Recall)}{(Precision + Recall)} \quad (3)$$

After completing the initial model training, the YOLO version demonstrating superior performance in terms of the F- score was selected for an additional round of semi-supervised learning, which leverages unlabeled data to further enhance the model's capabilities. To this end, a supplementary set of 200 images was procured from the original dataset, ensuring their distinction from those employed in the previous 5-fold training process. Considering we have different size YOLO models (n/s/m/l/x), during the semi-supervised learning phase, these newly obtained unlabeled images were fed into a selected large YOLO model with the best performance, producing output predictions that functioned as pseudo-labels for subsequent training iterations of a small YOLO model with faster inference. These pseudo-labels well served as surrogates and were amalgamated with the original labeled dataset, thereby augmenting the training dataset. Consequently, the model underwent further training using this expanded dataset, which now encompassed both labeled and pseudo-labeled instances. Here, the method that a more accurate model is used to generate the pseudo-labels for the unlabeled data, and then using these to train a smaller model, can be seen as a kind of "knowledge distillation", where the insights of the larger model are being transferred to the smaller model. We call it the semi-supervised learning for that both labeled and unlabeled (pseudo-labeled) data are used during training of the small model. By training on this combined dataset, models were encouraged to discover and capture latent features, relations, or properties present in the data without relying on external supervision. This approach allows models to learn in a more autonomous and data-driven manner, facilitating the development of representations that are useful for downstream tasks.

Following the semi-supervised learning phase, the most suitable model was selected and integrated into a pipeline with ByteTrack. Initially, the video was processed frame by frame using YOLO as the iPAK4 protein detector, generating bounding box predictions for iPAK4 in each frame. These bounding box predictions were then forwarded to ByteTrack for object tracking. ByteTrack employs an association algorithm to link the bounding boxes of the same object across consecutive frames and assigns a unique identifier (ID) to each object. This pipeline synergistically combined YOLO's accurate initial detections and ByteTrack's object associations, leading to refined tracking results. The resulting IDs and corresponding bounding box information for each iPAK4 in every frame of the input video were stored in a Python dictionary and saved as a npy file for subsequent post-tracking processing and analysis.

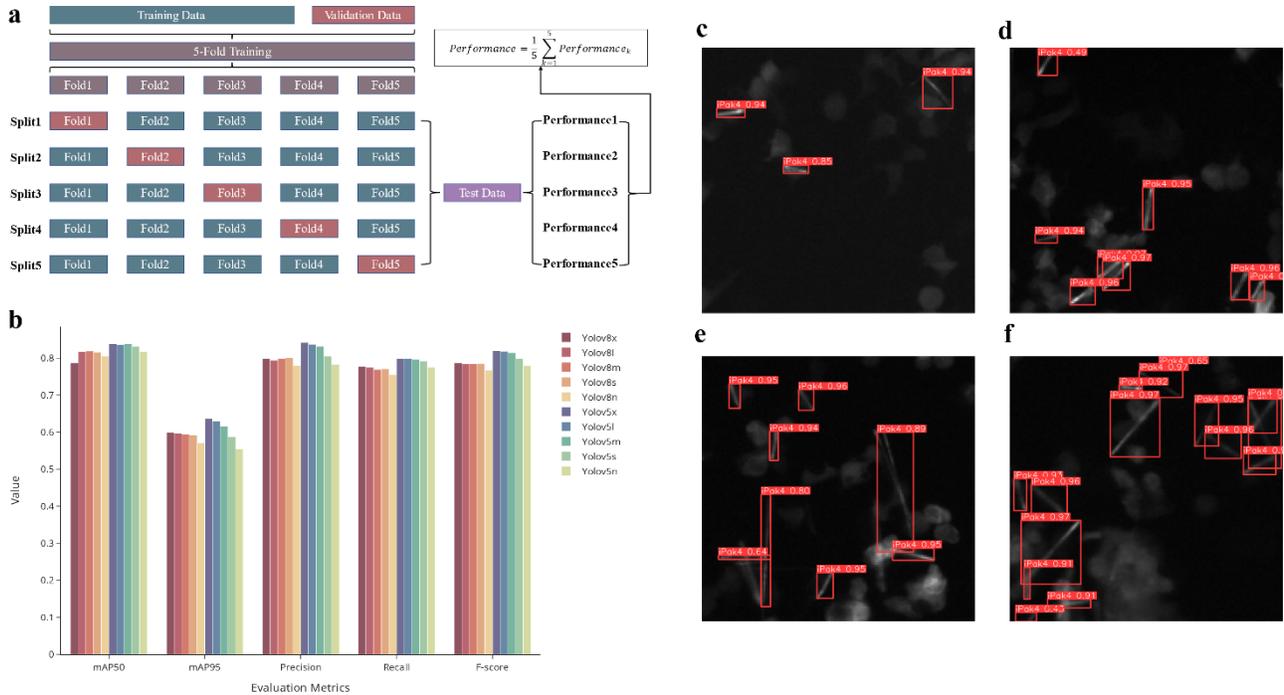

**Figure 2: a,** Schematic representation of the 5-fold cross-validation training process. **b,** Performance comparison of the models across five evaluation metrics. The results demonstrate the consistent superiority of YOLOv5 over YOLOv8 on the dataset. **c-f,** Some detection results from YOLOv5x model of iPAK4 protein fibers time-lapse videos at different time points.

*C. Post-tracking Processing*

The video generated from the previous steps served as the basis for manually selecting the IDs of accurately tracked iPAK4 proteins, and the bounding box coordinates for each ID across the video frames were then retrieved from the corresponding npy file.

To analyze the growth profile of the iPAK4 proteins, a region of interest (ROI) was cropped for each selected protein in its respective frame of the video. In order to accommodate the diverse orientations and offsets of the proteins from the ROI center, the Radon transform technique was utilized. By computing the Radon transform within the range of angles from 0° to 180° for each cropped ROI, a sinogram was generated. The peak of sinogram indicates the angle and offset where the integration of pixel intensities attained its maximum, thereby providing information about the precise location of the iPAK4 protein. The corresponding line in the real-space movie was then utilized to calculate the protein's length profile. To enhance accuracy, the intensities of 11 parallel lines on either side of the protein were averaged. Lastly, the ends of the protein were identified by applying a threshold to the intensity versus position plot, and the protein length at that specific timepoint was calculated using the Euclidian distance between the endpoint coordinates. The change in protein length across the video represents the growth profile of the target iPAK4 protein. Afterwards, the growth profiles of multiple iPAK4 proteins were pooled together and the group performance was analyzed.

III. RESULTS AND DISCUSSION

The results of this detection-based tracking for iPAK4 proteins are presented below. As mentioned in previous section, total 10 models of YOLOv8 and YOLOv5 were trained on the same dataset with 5-fold cross-validation methods, and their final performance in terms of mAP50, mAP95, precision, recall, F-score are compared in *Figure 2b* above.

Among the evaluated models, YOLOv5 consistently outperformed YOLOv8 across all metrics. YOLOv5x demonstrated the highest performance, achieving an mAP50 of 0.893, an mAP95 of 0.637, a precision of 0.842, recall of 0.798, and an F-score of 0.820. These results indicate that YOLOv5x exhibited superior object detection accuracy and achieved a better balance between precision and recall compared to all the other models. In contrast, the YOLOv8 models (for all n/s/m/l/x), showed relatively lower scores across all metrics. Although YOLOv8n exhibited the highest mAP50 among the YOLOv8 variants, it still fell short compared to the YOLOv5 models. Representative iPAK4 protein fibers detection results from the YOLOv5x are illustrated in *Figure 2c-f*.

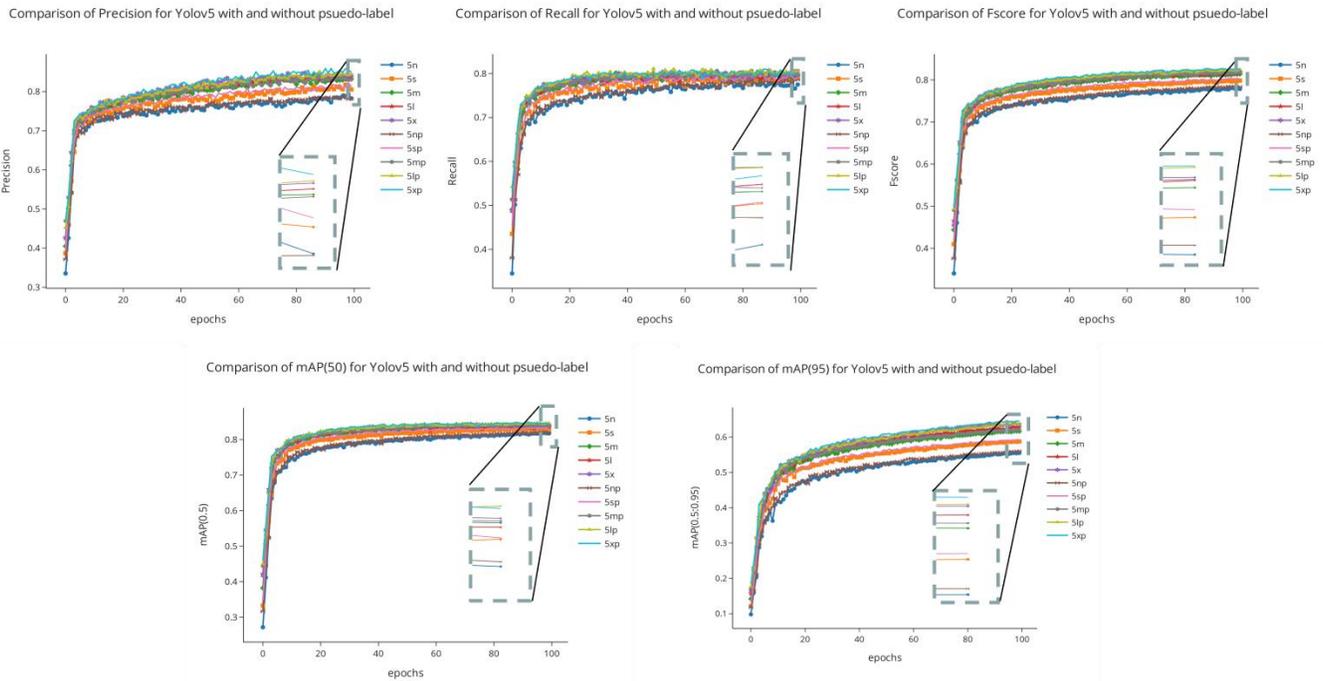

**Figure 3:** Performance Comparison before and after Semi-Supervised Learning. The legend includes a suffix "p" denoting the integration of pseudo-labels through semi-supervised learning. Notably, with an equal number of training epochs, the application of semi-supervised learning consistently enhances performance across all evaluation metrics.

The performance disparities observed between the YOLOv8 and YOLOv5 models in this study can be attributed to several factors. One potential contributing factor is the architectural differences between these models. While both YOLOv8 and YOLOv5 belong to the same model series and share the CSPDarknet53 backbone and a similar head structure, YOLOv8 introduces additional components such as a self-attention mechanism and a feature pyramid network in its head. These architectural modifications are intended to enhance the model's performance in long range contextual dependencies. However, it is crucial to note that the effectiveness of such modifications can vary depending on the characteristics of the dataset. In the case of our time-lapse dataset of iPAK4 protein, the introduction of self-attention mechanisms and feature pyramid network may not yield the same benefits as observed in general detection scenarios.

As YOLOv5 outperformed YOLOv8 in all metrics, they were then improved further with semi-supervised learning as mentioned in the methodology section. The effectiveness of semi- supervised training was then evaluated, and the results can be referred to *Figure 3*. It can be found that the incorporation of semi-supervised learning indeed resulted in performance improvements across various metrics for all the YOLOv5 models. Take Yolov5n which is the smallest model as an example, it achieved an F-score of 0.778 prior to the semi- supervised learning. However, after integration of the pseudo-labels training, its F-score escalated to 0.784. This trend of improvement was consistent across other YOLOv5 models as well, providing compelling evidence that semi-supervised learning is a highly effective strategy for leveraging unlabeled data to enhance the performance of object detectors. The results affirm that semi-supervised learning can further advance the performance of deep learning models while simultaneously reducing the labor-intensive nature of data annotation. After conducting a comprehensive evaluation and improvement of detection performance, the effectiveness of the complete detection-based tracking framework was assessed for capturing the dynamic behavior of iPAK4 protein fibers.

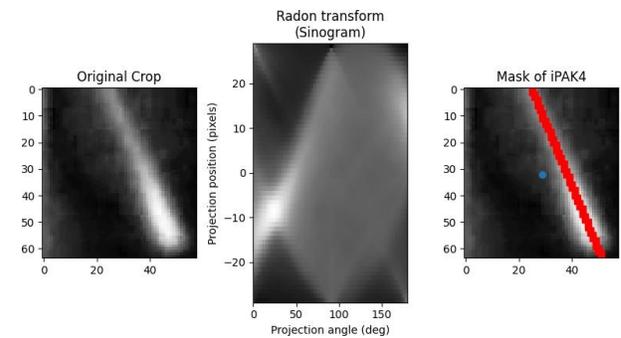

**Figure 4:** Radon Transform-Based Positioning of iPAK4 Protein Fibers. The peak of sinogram is associated with the exact position of iPAK4 fiber within the tracking ROI.

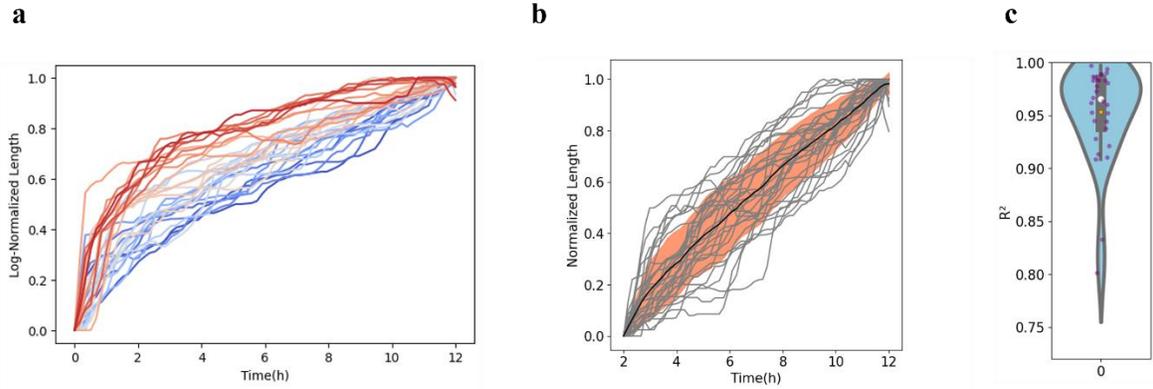

**Figure 5: a,** Growth profiles of iPAK4 fibers (n=31) captured using our proposed framework reveal two distinct phases: an initial rapid growth phase followed by a slower linear growth phase. **b,** Population average (black line) and standard deviation (orange line) of linear-phase fiber growth are depicted, with individual growth profiles shown as gray lines (n=31). **c,** Violin plot with embedded box-whisker plot illustrates the distribution of R2 values obtained from linear fitting.

Considering the computational constraints typically encountered in biology laboratories, we selected YOLOv5n as it requires minimal computational resource. Following the integration of semi-supervised learning, YOLOv5n was combined with the ByteTrack algorithm for this demonstration.

By utilizing input videos of iPAK4 time-lapse recordings, the growth activity of 31 individual protein fibers were tracked and their growth profiles over the designated time period were extracted, as shown in *Figure 5a*. The obtained plot distinctly reveals two phases in the growth behavior of iPAK4 fibers. Initially, a rapid exponential growth phase was observed during the first 1-2 hours, followed by a subsequent transition to a slower linear phase, as illustrated in *Figure 5b*. To evaluate the consistency and reliability of our framework, a linear regression analysis was performed on the population growth profiles of the 31 iPAK4 protein fibers. The results demonstrated a population average $R^2$ value of 0.95 ± 0.04 (mean ± s.d.; *Figure 5c*), affirming the strong agreement between the extracted growth profiles and the kinetic model previously reported in a study [19] (for detailed definition, please refer to its Supplementary Calculation 1):

$$L(t) = \frac{k_{synth}}{A\rho}t + \frac{V}{A\rho}(C_{nuc} - C_{ss})\left(1 - e^{-\frac{A\alpha k_{grow}}{V}t}\right) \quad (4)$$

## IV. CONCLUSION

In this work, we present a novel approach, *ProGroTrack*, for tracking cellular and biomolecular activities by combining the YOLO and ByteTrack algorithms within the DBT framework. Evaluating the performance of YOLOv5 and YOLOv8 models, we found that YOLOv5 outperformed YOLOv8 in all metrics, with YOLOv5x achieving the highest F-score. Incorporating semi-supervised learning further improved detection accuracy and reduced manual effort. As a representative case study, we successfully applied our approach to track the growth behavior of iPAK4 protein fibers over a designated time period. The analysis revealed two distinct growth phases that align with a previously reported kinetic model. The implications of our work extend beyond iPAK4 protein fibers, as our proposed method holds promise for automating the continuous tracking of experiments involving highly deformable objects, such as bacteria, cells, and other intracellular organelles. Future research can focus on consolidating the algorithms to ensure optimal object detection performance and minimize tracking losses.

ACKNOWLEDGMENT

This work is supported by NIH MIRA Award (1R35GM147274) and Johns Hopkins University.